\documentclass[final]{IEEEtran}

\IEEEoverridecommandlockouts

\usepackage[cmex10]{amsmath}
\usepackage{amsfonts}
\usepackage{algorithmic}
\usepackage{array}
\usepackage{mdwmath}
\usepackage{mdwtab}
\usepackage{eqparbox}
\usepackage[font=footnotesize]{subfig}
\usepackage{amsfonts}
\usepackage{tikz}
\usepackage{amsmath}
\usepackage{amsthm}
\usepackage{array}
\usepackage{mdwmath}
\usepackage{mdwtab}
\usepackage{eqparbox}
\usepackage{amsfonts}
\usepackage{tikz}
\usepackage{amssymb}
\usepackage{soul}

\newtheorem{remark}{Remark}

\newtheorem{theorem}{Theorem}

\newtheorem{lemma}{Lemma} 
\newtheorem*{lemma*}{Lemma} 
\newtheorem{corollary}{Corollary}
\newtheorem*{axiom*}{Data Processing Axiom}
\newtheorem*{question*}{Question}
\newtheorem{question}{Question}

\newtheorem{definition}{Definition}

\def \cX {\mathcal{X}}
\def \cR {\hat{\mathcal{X}}}

\def \bE {\mathbb{E}}
\def \bR {\mathbb{R}}

\def \logloss {\ell_{\mathrm{log}}}
\def \to {\rightarrow}

\begin{document}

\sloppy

\title{Justification of Logarithmic Loss via the Benefit of Side Information}


\author{Jiantao~Jiao,~\IEEEmembership{Student Member,~IEEE},~Thomas~Courtade,~\IEEEmembership{Member,~IEEE},~Kartik~Venkat,~\IEEEmembership{Student Member,~IEEE}, and Tsachy~Weissman,~\IEEEmembership{Fellow,~IEEE}

\thanks{Manuscript received Month 00, 0000; revised Month 00, 0000; accepted Month 00, 0000. Date of current version Month 00, 0000.
This work was supported in part by the Center for Science of Information (CSoI), an NSF Science and Technology Center, under grant agreement CCF-0939370. The material in this paper was presented in part at the 2014 IEEE International Symposium on
Information Theory, Honolulu, HI, USA. Copyright (c) 2014 IEEE. Personal use of this material is permitted. However, permission to use this material for any other purposes must be obtained from the IEEE by sending a request to pubs-permissions@ieee.org. 
}%

\thanks{Jiantao Jiao, Kartik Venkat, and Tsachy Weissman are with the Department of Electrical Engineering, Stanford University. Email: \{jiantao,kvenkat,tsachy\}@stanford.edu}
\thanks{Thomas Courtade is with the Department of Electrical Engineering and Computer Sciences, University of California, Berkeley. Email: {courtade@eecs.berkeley.edu}}

\thanks{This work was supported in part by the NSF Center for Science of Information under grant agreement CCF-0939370.}

}

\maketitle

\maketitle

\begin{abstract}

We consider a natural measure of relevance: the reduction in  optimal prediction risk in the presence of side information. For any given loss function, this relevance measure captures the benefit of side information for performing inference on a random variable under this loss function. When such a measure satisfies a natural data processing property, and the random variable of interest has alphabet size greater than two, we show that it is uniquely characterized by the mutual information, and the corresponding loss function coincides with logarithmic loss. In doing so, our work provides a new characterization of mutual information, and justifies its use as a measure of relevance. When the alphabet is binary, we characterize the only admissible forms the measure of relevance can assume while obeying the specified data processing property. Our results naturally extend to measuring causal influence between stochastic processes, where we unify different causal-inference measures in the literature as instantiations of directed information.
\end{abstract}





\section{Introduction}

In statistical decision theory, it is often a controversial issue to choose the appropriate loss function. One popular loss function is called \emph{logarithmic loss}, defined as follows. Let $\cX = \{x_1,x_2,\ldots,x_n\}$ be a finite set with $|\cX| = n$, let $\Gamma_n$ denote the set of probability measures on $\cX$, and let $\bar{\bR}$ denote the extended real line. 

\begin{definition}[Logarithmic Loss]\label{def.logloss}
Logarithmic loss  $\logloss \colon \cX \times \Gamma_n \to \bar{\mathbb{R}}$ is defined by 
\begin{equation} \label{eq.loglossdef}
\logloss(x,P) = \log \frac{1}{P(x)},
\end{equation}
where $P(x)$ denotes the probability of $x$ under measure $P$. 
\end{definition}

Logarithmic loss has enjoyed numerous applications in various fields. For instance, its usage in statistics dates back to Good \cite{Good1952}, and it has found a prominent role in learning and prediction (cf. Cesa--Bianchi and Lugosi \cite[Ch.\ 9]{Cesa--Lugosi2006}).
Logarithmic loss also assumes an important role in information theory, where many of the fundamental quantities (e.g., entropy, relative entropy, etc.) can be interpreted as the optimal prediction risk or regret under logarithmic loss. Recently, Courtade and Weissman~\cite{Courtade--Weissman2014} showed that the long-standing open problem of multiterminal source coding could be completely solved under logarithmic loss, providing yet another concrete example of its special nature. The use of the logarithm in defining entropy arises due to its various axiomatic characterizations, the first of which dates back to Shannon  \cite{Shannon1948}. 

%


The main contribution of this paper is in providing fundamental justification for inference using logarithmic loss. In particular, we show that a single and natural Data Processing requirement mandates the use of logarithmic loss. We begin by posing the following:

\begin{question}[Benefit of Side Information]\label{ques.main}
Suppose $X,Y$ are dependent random variables. How relevant is $Y$ for inference on $X$?  
\end{question}

\section{Problem Formulation and Main Results}

Toward answering Question~\ref{ques.main}, let $\ell\colon \cX \times \hat{\mathcal{X}} \to \bar{\bR}$ be an arbitrary loss function with reconstruction alphabet $\cR$, where $\cR$ is arbitrary. Given $(X,Y) \sim P_{XY}$,  it is natural to quantify the benefit of additional side information $Y$ by computing the difference between the expected losses in estimating $X \in \mathcal{X}$ with and without side information $Y$, respectively.  This motivates the following definition:
\begin{align}
C(\ell, P_{XY}) \triangleq \: \!\!\!\inf_{\hat{x}_1 \in \cR} \!\bE_P [\ell(X,\hat{x}_1)] -\! \inf_{\hat{X}_2(Y)} \bE_P [\ell(X,\hat{X}_2)],\label{eqn.overallben}
\end{align}
where $\hat{x}_1 \in \cR$ is deterministic, and $\hat{X}_2 = \hat{X}_2(Y) \in \cR$ is any measurable function of $Y$. In the following discussions, we require that indeterminate forms like $\infty-\infty$ do not appear in the definition of $C(\ell, P_{XY})$. By taking $Y$ to be independent of $X$, this requirement implies that for all $P \in \Gamma_n$, 
\begin{align}
\Big |\inf_{\hat{x}_1 \in \cR} \!\bE_P [\ell(X,\hat{x}_1)] \Big |<\infty. \label{eqn.vlessthaninfty}
\end{align}

The formulation (\ref{eqn.overallben}) has appeared previously in the statistics literature. DeGroot~\cite{degroot1962} in 1962 defined the \emph{information} contained in an experiment, which turns out to be equivalent to \eqref{eqn.overallben}. Later, Dawid~\cite{Dawid1998} defined the \emph{coherent dependence function}, which is equivalent to \eqref{eqn.overallben}, and used it to quantify the dependence between two random variables $X, Y$. Our framework of quantifying the predictive benefit of side information is closely connected to the notion of {proper scoring rules} and the literature on probability forecasting in statistics. The survey by Gneiting and Raftery \cite{Gneiting--Raftery2007} provides a good overview. 

Having introduced the yardstick in (\ref{eqn.overallben}), we now reformulate the question of interest: \textit{Which loss function(s) $\ell$ can be used to define $C(\ell, P_{XY})$ in a meaningful way?} Of course, ``meaningful" is open to interpretation, but it is desirable that $C(\ell, P_{XY})$ be well-defined, at minimum.  This motivates the following axiom:

\begin{axiom*}
For all distributions $P_{XY}$, the quantity $C(\ell, P_{XY})$ satisfies
\begin{equation}
C(\ell, P_{TY}) \leq C(\ell, P_{XY}) \notag
\end{equation}
whenever $T(X) \in \cX$ is a statistically sufficient transformation of $X$ for $Y$.
\end{axiom*}
We remind the reader that  the statement `$T$ is a statistically sufficient transform of $X$ for $Y$' means that the following two Markov chains hold:
\begin{equation}
T - X - Y,\quad X - T - Y
\end{equation} 
That is, $T(X)$ preserves all of the information $X$ contains about $Y$. 

In words, the Data Processing Axiom stipulates that processing the data $X \to T$ cannot boost the predictive benefit of the side information\footnote{In fact, the Data Processing Axiom is weaker than this general data processing statement since it only addresses statistically sufficient transformations of $X$.}.


To convince the reader that the Data Processing Axiom is a natural requirement, suppose instead that the Data Processing Axiom did not hold.  Since $X$ and $T$ are mutually sufficient statistics for $Y$, this would imply that there is \emph{no} unique value which quantifies the benefit of side information $Y$ for the random variable of interest. Thus, the Data Processing Axiom is needed for the benefit of side information to be well-defined.

Although the Data Processing Axiom may seem to be a benign requirement, it has far-reaching implications for the form $C(\ell, P_{XY})$ can take.  This is captured by our first main result:
\begin{theorem}\label{thm.main}
Let $n\geq 3$.  Under the Data Processing Axiom, the function $C(\ell, P_{XY})$ is uniquely determined by the mutual information,
\begin{equation}
C(\ell, P_{XY}) = I(X; Y), 
\end{equation}
up to a multiplicative factor.
\end{theorem}

The following corollary immediately follows from Theorem~\ref{thm.main}. 
\begin{corollary}\label{cor.add}
Let $n\geq 3$. Under the Data Processing Axiom, the benefit of additional side information $Y$ for inference on $X$ with common side information $W$, i.e.
\begin{equation}
\inf_{\hat{X}_1(W)} \bE_P [\ell(X,\hat{X}_1)] - \inf_{\hat{X}_2(Y,W)} \bE_P [\ell(X,\hat{X}_2)],
\end{equation}
is uniquely determined by the conditional mutual information,
\begin{equation}
I(X;Y|W),
\end{equation}
up to a multiplicative factor.
\end{corollary}
Thus, up to a multiplicative factor, we see that logarithmic loss generates the \emph{only} measure of predictive benefit (defined according to \eqref{eqn.overallben}) which satisfies the Data Processing Axiom.  In other words, Theorem \ref{thm.main} provides a definitive answer to Question 1 under the framework we have described, and also highlights the special role played by logarithmic loss.

Theorem~\ref{thm.main} shows that mutual information uniquely quantifies the reduction of prediction risk due to side information. Note that the characterization of mutual information afforded by Theorem~\ref{thm.main} does not explicitly require any of the mathematical properties of mutual information, such as the chain rule, or invariance to one-to-one transformations of both $X$ and $Y$. Thus, beyond the
operational implications of our result,  Theorem~\ref{thm.main} has strong implications for axiomatic characterization of information measures from a mathematical standpoint. On this point, we note that Csisz{\'a}r, in his survey~\cite{Csiszar2008} names the axiomatic result of Ac{\'z}el, Forte, and Ng \cite{Aczel--Forte--Ng1974} as ``\emph{intuitively most appealing}'' in characterizing the entropy in terms of symmetry, expansibility, additivity, and subadditivity, whereas most other known characterizations require recursivity or the sum property. For details we refer to Csisz{\'a}r~\cite{Csiszar2008}. 

Theorem~\ref{thm.main} provides a partial explanation for why mutual information is widely used as an inferential tool across various applications in science and engineering, and is deeply imbued in fundamental concepts in various disciplines. In statistics, one of the popular criteria for objective Bayesian modeling~\cite{Lehmann--Casella1998theory} is to design a prior on the parameter to maximize the mutual information between the parameter and the observations. In machine learning, the so-called \emph{infomax}~\cite{Linsker1988self} criterion states that the function that maps a set of input values to a set of output values should be chosen or learned so as to maximize the mutual information between the input and output, subject to a set of specified constraints. This principle has been widely adopted in practice, for example, in decision tree based algorithms in machine learning such as C4.5~\cite{Quinlan1993c4}, one tries to select the feature at each step of tree splitting to maximize the mutual information (called \emph{information gain} principle~\cite{Nowozin2012improved}) between the output and the feature conditioned on previous chosen features. In some applications, mutual information arises naturally as the only answer, for example, the well known Chow--Liu algorithm \cite{Chow--Liu1968} for learning tree graphical models relies on estimation of the mutual information, which is a natural consequence of maximum likelihood estimation. We also mention genetics~\cite{Olsen--Meyer--Bontempi2009impact}, image processing~\cite{Pluim--Maintz--Viergever2003mutual}, computer vision~\cite{Viola--Wells1997alignment}, secrecy~\cite{Batina--Gierlichs--Prouff--Rivain--Standaert--Veyrat2011mutual}, ecology~\cite{Hill1973diversity}, and physics~\cite{Franchini--Its--Korepin2008Renyi} as fields in which mutual information is widely used. Erkip and Cover~\cite{erkip1998efficiency} argued that mutual information was a natural quantity in the context of portfolio theory, where it emerges as the increase in growth rate due to the presence of side information. 


Mutual information and related information theoretic measures are instrumental in various applications. This motivates investigating optimal estimators for these quantities based on data. There exist extensive literature on this subject, and we refer to~\cite{Jiao--Venkat--Han--Weissman2014minimax} for a detailed review, as well as the theory and Matlab/Python implementations of entropy and mutual information estimators that achieve the minimax rates in all the regimes of sample size and support size pairs. For the recent growing literature on information measure estimation in the high-dimensional regime, we refer to~\cite{Valiant--Valiant2011, Valiant--Valiant2011power,Valiant--Valiant2013estimating, Jiao--Venkat--Han--Weissman2014minimax, Wu--Yang2014minimax, Jiao--Venkat--Han--Weissman2014MLE, Acharya--Orlitsky--Suresh--Tyagi2014complexity, Jiao--Venkat--Han--Weissman2014beyond}. 

Interestingly, the assumption that $n\geq 3$ in Theorem \ref{thm.main} is essential. When the alphabet of $X$ is binary, i.e. $n=2$, the Data Processing Axiom no longer mandates the use of logarithmic loss. We have an explicit characterization for the form the measure of relevance~(\ref{eqn.overallben}) can take. The class of solutions for the binary alphabet setting  is characterized by the following theorem.

\begin{theorem}\label{thm.main2}
Let  $n = 2$.  Under the Data Processing Axiom, $C(\ell, P_{XY})$ must be of the form
\begin{equation}\label{eqn.defineg2new}
C(\ell, P_{XY})  = \sum_{y} P_Y(y) G(P_{X|Y=y}) - G(P_{X}),
\end{equation}
where $G((p,1-p)):\Gamma_2 \to \bR$ is a symmetric (invariant to permutations), convex function. Moreover, for any symmetric convex function $G((p,1-p)):\Gamma_2 \to \bR$, there exists a loss function $\ell$ whose corresponding $C(\ell, P_{XY})$ satisfies the Data Processing Axiom and is given by~\eqref{eqn.defineg2new}.

\end{theorem}

It is worth mentioning that there is an interesting set of observations surrounding the characterization of information measures which is sensitive to the alphabet size being binary or larger. This phenomenon is explored further in \cite{Jiao--Courtade--No--Venkat--Weissman2014}. 

The rest of this paper is organized as follows. In Section \ref{sec:Causality}, we explore the connections between our results and the existing literature on causal analysis, including Granger and Sims causality, Geweke's measure, transfer entropy, and directed information. The proofs of Theorems~\ref{thm.main} and \ref{thm.main2} are provided in Section \ref{sec:Proofs}. Proofs of some auxiliary lemmas are deferred to the appendix.   

\section{Causality Measures: An Axiomatic Viewpoint}\label{sec:Causality}

Inferring causal relationships from observed data plays an indispensable part in  scientific discovery. Granger, in his seminal work \cite{Granger1969},  proposed a predictive  test for inferring causal relationships. To state his test, let $X_t,Y_t,U_t$ be stochastic processes, where $X_t,Y_t$ are the processes of interest, and $U_t$ contains all information in the universe accumulated up to time $t$. Granger's causality test asserts that $Y_t$ causes $X_t$, denoted by $Y_t\Rightarrow X_t$, if we are better able to predict $X_t$ using the past information of $U_t$, than by using all past information in $U_t$ {apart from} $Y_t$.  In Granger's definition, the quality of prediction is measured by the squared error  risk achieved by the optimal unbiased least-squares predictor.
%
%

In his 1980 paper, Granger \cite{Granger1980} introduced a set of operational definitions which made it possible to derive practical testing procedures. For example, he assumes that we must be able to specify $U_t$ in order to perform causality tests, which is slightly different from his original definition which required knowledge of all information in the universe (which is usually unavailable). 

Later, Sims \cite{Sims1972} introduced a related concept of causality, which was  proved to be equivalent to Granger's definition in Sims \cite{Sims1972}, Hosoya \cite{Hosoya1977}, and Chamberlain \cite{Chamberlain1982} in a variety of settings. 

Motivated by Granger's  framework for testing causality using linear prediction, Geweke \cite{Geweke1982}\cite{Geweke1984} proposed a {causality measure} to quantify the extent to which $Y$ is causing $X$. Quoting Geweke (emphasis ours):
\vspace{.5ex}

\begin{center}
\parbox{.45\textwidth}{~~\emph{ \small ``The empirical literature abounds with \underline{tests} of independence and unidirectional causality for various pairs of time series, but there have been virtually no investigations of the \underline{degree} of dependence or the extent of various kinds of feedback. The latter approach is more realistic in the typical case in which the hypothesis of independence of unidirectional causality is not literally entertained, but it requires that one be able to measure linear dependence and feedback."}}
\end{center}
\vspace{1.5ex}

\noindent In other words, Geweke makes the important distinction between a \emph{causality test} which makes a binary decision on whether one process causes another, and a \emph{causality measure} which quantifies the degree to which one process causes another.  Geweke proposed the following measure as a natural starting point:
\begin{equation}
F_{Y \Rightarrow X} \triangleq \ln \frac{\sigma^2 (X_t|X^{t-1})}{\sigma^2(X_t|X^{t-1},Y^{t-1})}, \label{gewekeMeasure}
\end{equation}
where $\sigma^2(X_t|X^{t-1},Y^{t-1})$ is the variance of the prediction residue when predicting $X_t$ via the optimal linear predictor constructed from observation $X^{t-1}, Y^{t-1}$. Note that if $F_{Y \Rightarrow X} > 0$, we could conclude $Y_t \Rightarrow X_t$ according to Granger's test. 

It has long been observed that the restriction to optimal linear predictors in testing causality is not necessary. In fact, Chamberlain \cite{Chamberlain1982} proved a general equivalence between  Granger and Sims' causality tests by replacing linear predictors with conditional independence tests. However, the natural generalization of  \eqref{gewekeMeasure} wasn't clear until Gourieroux, Monfort, and Renault \cite{Gourieroux--Monfort--Renault1987} proposed the so-called \emph{Kullback causality measures} in 1987. It is now well-known that  Kullback causality measures are equivalent to \eqref{gewekeMeasure} under linear Gaussian models (e.g. Barnett, Barrett and Seth \cite{Barnett--Barrett--Seth2009}). 

Using information theoretic terms,  Kullback causality measures are nothing but the directed information introduced by Massey \cite{Massey90}, and motivated by Marko \cite{Marko1973}. Using modern notation, the directed information from $X^n$ to $Y^n$ is defined as
\begin{align}
I(X^n \to Y^n) & \triangleq \sum_{i = 1}^n I(X^i; Y_i|Y^{i-1}) \\
& = H(Y^n) - H(Y^n \| X^n),
\end{align}
where $H(Y^n\| X^n)$ is the \emph{causally conditional entropy}, defined by  
\begin{equation}
H(Y^n \| X^n) \triangleq \sum_{i = 1}^n H(Y_i | Y^{i-1}, X^i). 
\end{equation}

Massey and Massey \cite{Massey--Massey2005} established the pleasing conservation law of directed information:
\begin{align}
I(X^n;Y^n) & = I(X^n \to Y^n) + I(Y^{n-1} \to X^n) \\
& = I(X^{n-1} \to Y^n) + I(Y^{n-1} \to X^n) \nonumber \\
& \quad + \sum_{i = 1}^n I(X_i;Y_i|X^{i-1}, Y^{i-1}), 
\end{align}
which  implies that the extent to which process $X_t$ influences process $Y_t$ and vice-versa always sum to the total mutual information between the two processes. 
Since $I(Y^{n-1}\to X^n)$ can be expressed as
\begin{equation}
I(Y^{n-1}\!\to X^n) = \sum_{i = 1}^n H(X_i|X^{i-1}) - H(X_i|X^{i-1},Y^{i-1}),\notag
\end{equation}
$X_i$ being conditionally independent of $Y^{i-1}$ given $X^{i-1}$ is equivalent to $I(Y^{n-1}\to X^n) = 0$.  This corresponds precisely to the definition of general Granger non-causality. Permuter, Kim, and Weissman~\cite{Permuter--Kim--Weissman2011} showed various applications of directed information in portfolio theory, data compression, and hypothesis testing in the presence of causality constraints. Amblard and Michel~\cite{Amblard--Michel2012} reviewed the intimate connections between Granger causality and directed information theory. 

We remark that, for practical applications, the directed information between stochastic processes can be computed using the universal estimators proposed in \cite{Jiao--Permuter--Zhao--Kim--Weissman2013}, which exhibit near-optimal statistical properties. 

Finally, we note that the notion of \emph{transfer entropy} in the physics literature, which was proposed by Schreiber \cite{Schreiber2000} in 2000,  turns out to be equivalent to directed information. 

To connect our present discussion on causality measures to Theorem \ref{thm.main}, we recall that the directed information rate \cite{Kramer1998} between a pair of jointly stationary finite-alphabet processes $X_t,Y_t$ can be written as:
\begin{align}
& \lim_{n\to \infty}\frac{1}{n} I(Y^{n-1}\to X^n)\nonumber \\
&~~= \inf_{ T_1(X^{-1}_{-\infty})} \bE[\logloss(X_0, T_1)] ~~- \!\!\!\!\inf_{T_2(X^{-1}_{-\infty},Y^{-1}_{-\infty})} \!\!\bE[\logloss(X_0,T_2)]. \notag
\end{align}

In light of this, we can conclude from  Theorem~\ref{thm.main} and Corollary~\ref{cor.add} that the directed information rate is the \emph{unique} measure of causality which assumes the form \eqref{eqn.overallben} and satisfies the Data Processing Axiom. Thus, our axiomatic viewpoint  explains why the same causality measure has appeared so often in varied fields  including economics, statistics, information theory, and physics. Except in the binary case, we roughly have the following: \emph{All `reasonable' 
causality measures defined by a difference of predictive risks must coincide}.\footnote{Here, the authors' interpretation of ``reasonable" is reflected by the Data Processing Axiom.  In the context of this section, the Data Processing Axiom stipulates that any reasonable causality measure should be invariant under statistically sufficient transformations of the data -- a desirable property and natural criterion.}

\vspace{6pt}

\section{Proof of Main Results}\label{sec:Proofs}

In this section, we provide complete proofs of Theorems~\ref{thm.main} and \ref{thm.main2} and highlight the key ideas. 

To begin, we show that the measure of relevance defined in \eqref{eqn.overallben} is equivalently characterized by a bounded convex function defined on the $\mathcal{X}$-simplex. The following lemma achieves this goal. 

\begin{lemma}\label{lemma.jiao}
There exists a bounded convex function $V: \Gamma_n \to \bR$, depending on $\ell$, such that
\begin{align}
C(\ell, P_{XY}) = \left( \sum_{y} P_{Y}(y) V(P_{X|Y=y})\right) - V(P_{X}).
\end{align}
\end{lemma}
\noindent The proof of Lemma~\ref{lemma.jiao} follows from defining $V(P)$ by
\begin{equation}
V(P) = - \inf_{\hat{x} \in \cR}\bE_P [\ell(X,\hat{x})],
\end{equation}
and its details are deferred to the appendix. In the statistics literature, the quantity $-V(P)$ is usually called the \emph{generalized entropy} or the \emph{Bayes envelope}.  We refer to Dawid \cite{Dawid2007} for details. 

In the literature of concentration inequalities, the following functional
\begin{equation}
H_\Phi(Z) = \bE \Phi(Z) - \Phi(\bE Z),
\end{equation}
where $\Phi$ is a convex function, is called $\Phi$-entropy, which in fact quantifies the gap in Jensen's inequality. As shown in Lemma~\ref{lemma.jiao}, functional $C(\ell, P_{XY})$ is closely related to the notion of $\Phi$-entropy. We refer to Boucheron, Lugosi, and Massart \cite[Ch.\ 14]{Boucheron--Lugosi--Massart2013} for a nice survey on the usage of $\Phi$-entropies in proving concentration inequalities.

The next lemma asserts that we only need to consider symmetric (invariant to permutations) functions $V(P)$. 

\begin{lemma}\label{lemma.symmetric}
Under the Data Processing Axiom, there exists a symmetric finite convex function $G: \Gamma_n \to \bR$, such that
\begin{align}
C(\ell, P_{XY})= \left( \sum_{y} P_{Y}(y) G(P_{X|Y=y})\right) - G(P_{X}),
\end{align}
and $G(\cdot)$ is equal to $V(\cdot)$ in Lemma~\ref{lemma.jiao} up to a linear translation:
\begin{equation}
G(P) = V(P) + \langle c, P \rangle,
\end{equation}
where $c \in \bR^n$ is a constant vector.
\end{lemma}

The proof of Lemma~\ref{lemma.symmetric} follows by applying a permutation to the space $\cX$ and applying the Data Processing Axiom. Details are deferred to the appendix. 

Now we are in a position to begin the proof of Theorem~\ref{thm.main} in earnest. 

\subsection{The case $n\geq 3$}

It suffices to consider the case when the side information $Y$ is binary valued, i.e., $Y \in \{1,2\}$. We will show that the Data Processing Axiom mandates the usage of the logarithmic loss even when we constrain ourselves to this situation. 

Define $\alpha \triangleq \mathbb{P}\{Y=1\}$. Take $P_{\lambda_1}^{(t)}, P_{\lambda_2}^{(t)}$ to be two probability distribution on $\mathcal{X}$ parametrized in the following way:
\begin{align}
P_{\lambda_1}^{(t)} &  =(\lambda_1 t, \lambda_1 (1-t), r- \lambda_1  , p_4, \ldots ,p_n) \\
P_{\lambda_2}^{(t)} & = (\lambda_2 t,\lambda_2  (1-t), r-\lambda_2 , p_4, \ldots,p_n),
\end{align}
where $r \triangleq 1- \sum_{i\geq 4} p_i, t\in [0,1], 0 \leq \lambda_1 < \lambda_2 \leq r$. 

Taking $P_{X|1} \triangleq P_{\lambda_1}^{(t)}, P_{X|2} \triangleq P_{\lambda_2}^{(t)}$, it follows from Lemma~\ref{lemma.jiao} that
\begin{align}
&  C(\ell, P_{XY}) \nonumber \\
&\quad = \alpha V(P_{\lambda_1}^{(t)}) + (1-\alpha )V(P_{\lambda_2}^{(t)}) - V(\alpha P_{\lambda_1}^{(t)} + (1-\alpha )P_{\lambda_2}^{(t)}). 
\end{align}
 
Note that the following transformation $T(X)$ is a statistically sufficient transformation of $X$ for $Y$:
\begin{equation}
T(X) = \begin{cases} x_1 & X \in \{x_1,x_2\}, \\ X & \textrm{otherwise.}\end{cases}
\end{equation}

The Data Processing Axiom implies that for all $\alpha \in [0,1]$, $t \in [0, 1]$ and legitimate $\lambda_2 > \lambda_1 \geq 0$,
\begin{align}
& \alpha V(P_{\lambda_1}^{(t)}) + (1-\alpha )V(P_{\lambda_2}^{(t)}) - V(\alpha P_{\lambda_1}^{(t)} + (1-\alpha )P_{\lambda_2}^{(t)}) \nonumber \\
& \quad  = \alpha V(P_{\lambda_1}^{(1)}) + (1-\alpha )V(P_{\lambda_2}^{(1)}) - V(\alpha P_{\lambda_1}^{(1)} + (1-\alpha )P_{\lambda_2}^{(1)}). 
\end{align}

We now define the function
\begin{equation}
R(\lambda,t) \triangleq V(P_{\lambda}^{(t)}),
\end{equation}
where we note that the bi-variate function $R(\lambda, t)$ implicitly depends on the parameter $p_4, p_5 \ldots p_n$ which we shall fix for the rest of this proof. Thus, $R(\lambda,t) = R(\lambda,t; p_4,p_5,\ldots,p_n) $.

Note that by definition,
\begin{equation}
R(\alpha \lambda_1 + (1-\alpha) \lambda_2, t) = V(\alpha P_{\lambda_1}^{(t)} + (1-\alpha )P_{\lambda_2}^{(t)}),
\end{equation}
hence we know that
\begin{align}
& \alpha R(\lambda_1,t) + (1-\alpha)R(\lambda_2,t) -R(\alpha \lambda_1 + (1-\alpha) \lambda_2, t) \nonumber \\
& \quad = \alpha R(\lambda_1,1) + (1-\alpha)R(\lambda_2,1) -R(\alpha \lambda_1 + (1-\alpha) \lambda_2, 1). \label{eqn.abtoplug}
\end{align}

Taking $\lambda_1 = 0, \lambda_2 = r = 1-\sum_{i\geq 4}p_i$. We define $ \tilde{R}(\lambda,t) \triangleq R(\lambda,t)-\lambda U(t)$, where 
\begin{align}
U(t) = \frac{R(r,t)}{r}. 
\end{align}

It follows that 
\begin{equation}\label{eqn.abdesing}
\tilde{R}(0,t) = V(P_0^{(t)}), ~~~~\tilde{R}(r,t) = 0, ~~~\forall t\in [0,1],
\end{equation}
and we note that $V(P_0^{(t)})$ in fact does not depend on $t$. 

%
%
%
%
%
%
%

With the help of (\ref{eqn.abdesing}), we plug $R(\lambda,t) = \tilde{R}(\lambda,t) + \lambda U(t)$ into (\ref{eqn.abtoplug}), and obtain
\begin{equation}
\tilde{R}((1-\alpha)r,t) = \tilde{R}((1-\alpha)r, 1), ~~~~ \forall \alpha \in [0,1], t\in [0,1].
\end{equation}
In other words, there exists a function $E: [0,1] \to \bR$, such that 
\begin{equation}
\tilde{R}(\lambda,t) = E(\lambda).
\end{equation}

Since $R(\lambda,t) = \tilde{R}(\lambda,t) + \lambda U(t)$, we know that there exist real-valued functions $E, U$ (indexed by $p_4,\ldots,p_n$) such that
\begin{equation} \label{eqn.RUF}
R(\lambda,t) = \lambda U(t) + E(\lambda).
\end{equation}

Expressing $\lambda,t$ in terms of $p_1,p_2$, we have
\begin{equation}
\lambda = p_1 + p_2, \quad t = \frac{p_1}{p_1 + p_2}. 
\end{equation}
By definition of $R(\lambda,t)$, we can re-write (\ref{eqn.RUF}) as
\begin{align}
& V(p_1,p_2, p_3, p_4,\ldots,p_n)    \nonumber \\
&= (p_1 + p_2) U \left( \frac{p_1}{p_1 + p_2} ;  p_4,\ldots,p_n \right) \nonumber \\
& \quad    + E (p_1 + p_2 ;  p_4,\ldots,p_n).
\end{align}

By Lemma~\ref{lemma.symmetric}, we know that there exists a symmetric (permutation invariant) finite convex function $G: \Gamma_n \to \bR$, such that
\begin{equation}
G(P) = V(P) + \langle c, P \rangle.
\end{equation}
%
%
%
%

In other words, we have proved that $G$ is of the form
\begin{align}
& G(P) = (p_1 + p_2) U \left( \frac{p_1}{p_1 + p_2} ;  p_4,\ldots,p_n \right) \nonumber \\
& \qquad \qquad + E (p_1 + p_2 ;  p_4,\ldots,p_n) + \langle c, P \rangle.
\end{align}

For notational simplicity, we define
\begin{equation}
Y(p_1,p_2) \triangleq G(P),
\end{equation}
where we again note that $Y(p_1,p_2; p_4,\ldots,p_n)$ is a bi-variate function parameterized by $p_4, \ldots p_n$.  This gives
\begin{align}
Y(p_1,p_2) &  = (p_1 + p_2) U \left( \frac{p_1}{p_1 + p_2} \right) + E (p_1 + p_2) \nonumber \\
& \qquad + c_1 p_1 + c_2 p_2 + c_3 (r-p_1 - p_2).\label{eqn.defy}
\end{align}

Since $G(P)$ is a symmetric function, we know that if we exchange $p_1$ and $p_3$ in $G(P)$, the value of $G(P)$ will not change. In other words, for $r = p_1 + p_2 + p_3$, we have
\begin{align}
& (r-p_3) U \left( \frac{p_1}{r-p_3} \right) + E(r-p_3) + c_1 p_1 + c_2 p_2 + c_3 p_3 \nonumber \\
& \   = (r-p_1) U \left( \frac{p_3}{r-p_1} \right) + E(r-p_1) + c_1 p_3 + c_2 p_2 + c_3 p_1,
\end{align}
which is equivalent to
\begin{align}
& (r-p_3) U \left( \frac{p_1}{r-p_3} \right) + E(r-p_3) + (c_3 - c_1)p_3   \nonumber \\
& \  = (r-p_1) U \left( \frac{p_3}{r-p_1} \right) + E(r-p_1) + (c_3 - c_1)p_1.
\end{align}

Defining $\tilde{E}(x) \triangleq  E(r-x) +(c_3 - c_1)x$, we have
\begin{equation}
(r-p_3) U \left( \frac{p_1}{r-p_3} \right) + \tilde{E}(p_3)= (r-p_1) U \left( \frac{p_3}{r-p_1} \right) + \tilde{E}(p_1).
\end{equation}

Interestingly, we can solve for general solutions of the above functional equation, which has connections to the so-called \emph{fundamental equation of information theory}:

\begin{lemma*}[\cite{Kannappan--Ng1973}\cite{Maksa1982}\cite{Aczel--Ng1983}]
The most general measurable solution of
\begin{equation}
f(x) + (1-x) g \left( \frac{y}{1-x} \right) = h(y) + (1-y) k \left( \frac{x}{1-y} \right), \label{gensoln}
\end{equation}
for $x,y\in [0,1)$ with $x+y \in [0,1]$, where $f,h: [0,1) \to \bR$ and $g,k: [0,1] \to \bR$, has the form
\begin{align}
f(x) & = a H_2(x) + b_1 x + d,\label{fGenSoln} \\
g(y) & = a H_2(y) + b_2 y + b_1 - b_4, \\
h(x) & = a H_2(x) + b_3 x + b_1 + b_2 - b_3 - b_4 + d, \\
k(y) & = a H_2(y) + b_4 y + b_3 - b_2, \label{kGenSoln}
\end{align} 
for $x\in [0,1), y\in [0,1]$, where $H_2(x) = -x \ln x -(1-x) \ln (1-x)$ is the binary Shannon entropy and $a,b_1,b_2,b_3,b_4$, and $d$ are arbitrary constants.
\end{lemma*}

\begin{remark}
If $f = g = h = k$ in \eqref{fGenSoln}-\eqref{kGenSoln}, the corresponding functional equation is called the `{fundamental equation of information theory}'. 
\end{remark}

In order to apply the above lemma to our setting, we define
\begin{equation}
q_i = p_i/r, \quad i = 1,2,3
\end{equation}
and $h(x) = \tilde{E}(r x)/r$. Then we know
\begin{equation}
(1-q_3) U \left( \frac{q_1}{1-q_3} \right) + h (q_3) = (1-p_1) U \left( \frac{q_3 }{1-q_1}\right) + h(q_1). 
\end{equation}

Applying the general solution of \eqref{gensoln}, setting $f = h, g = k = U$, we have
\begin{equation}
b_1 = b_3, b_2 = b_4.
\end{equation}
Thus,
\begin{equation}
h(x) = a H_2(x) + b_1 x + d,
\end{equation}
\begin{equation}
U(y) = a H_2(y) + b_2 y + b_1 - b_2.
\end{equation}

By the definition of $h(x)$ and $\tilde{E}(x)$, we have that
\begin{equation}
E(x) = r a H_2(x/r) + ( b_1 + c_1 - c_3)(r-x) +d.
\end{equation}

Plugging the general solutions to $U(x), E(x)$ into (\ref{eqn.defy}), and redefining the constants, we have
\begin{align}
& Y(p_1,p_2) \nonumber \\
&  = A \Big ( p_1 \ln p_1 + p_2 \ln p_2 + (r-p_1 - p_2)\ln (r-p_1 - p_2) \Big ) \nonumber \\
& \qquad + Bp_1 + C p_2 + D.
\end{align}

Note that the constants $A,B,C,D$ are functions of $p_4,\ldots,p_n$. Therefore, we have the following general representation of the symmetric function $G(P)$:
\begin{align}
& G(P) = A(p_4,\ldots,p_n) \left( p_1 \ln p_1 + p_2 \ln p_2 + p_3\ln p_3 \right) \nonumber \\
& \qquad \qquad + B(p_4,\ldots,p_n)p_1 + C(p_4,\ldots,p_n)p_2 \nonumber \\
& \qquad \qquad +   D(p_4,\ldots,p_n), 
\end{align}
 where we have made the dependence on $p_4 \ldots p_n$ explicit. Now we utilize the property that $Y(p_1,p_2)$ is invariant to permutations. Exchanging $p_1,p_2$, we obtain that $B \equiv C$. Exchanging $p_1,p_3$, we obtain that $B \equiv C \equiv 0$. Doing an arbitrary permutation on $p_4,\ldots,p_n$, since $p_1,p_2,p_3$ enjoy two degrees of freedom, we know that $A(p_4,\ldots,p_n), D(p_4,\ldots,p_n)$ are symmetric functions.

Exchanging $p_1,p_4$ and comparing the coefficients for $p_2\ln p_2$, we know that 
\begin{equation}
A(p_4,p_5,\ldots,p_n) = A(p_1,p_5, \ldots,p_n) ,
\end{equation}
since $A$ is symmetric, and thus we can conclude that $A$ is a constant. Now exchanging $p_1,p_4$ gives us
\begin{equation}\label{eqn.dede}
A p_1 \ln p_1 - A p_4 \ln p_4 = D(p_1,p_5,\ldots,p_n) - D(p_4,p_5,\ldots,p_n).
\end{equation}

Taking partial derivatives with respect to $p_1$ (we vary $p_2$ simultaneously to ensure $P$ still lies on the simplex) on both sides of (\ref{eqn.dede}), we obtain
\begin{equation}
A \left( \ln p_1 + 1\right) = \frac{\partial}{\partial p_1} D(p_1,p_5,\ldots,p_n).
\end{equation}

Integrating on both sides with respect to $p_1$, we know there exists a function $f$ such that
\begin{equation}
D(p_1, p_5,\ldots,p_n) = A p_1 \ln p_1 + f(p_5,\ldots,p_n).
\end{equation}
Since $D$ is symmetric, we further know that
\begin{equation}
D(p_4,\ldots,p_n) = \sum_{i\geq 4} A p_i \ln p_i.
\end{equation}

To sum up, we have
\begin{equation}\label{eqn.gsolution}
G(P) = A \sum_{i = 1}^n p_i \ln p_i.
\end{equation}
To guarantee that $G(P)$ is convex, we need $A>0$. 

Plugging (\ref{eqn.gsolution}) into Lemma~\ref{lemma.symmetric}, the proof is complete.

\subsection{The case $n = 2$}
Under the Data Processing Axiom, Lemma~\ref{lemma.symmetric} implies the corresponding representation. On the other hand, for an arbitrary convex function $G$, the Savage representation of proper scoring rules \cite{Gneiting--Raftery2007} gives the construction of the corresponding loss function $\ell$. Indeed, the Savage representation asserts, for a convex function $G$, we can define a loss function $\ell_G(x,Q): \cX \times \Gamma_n \to \bar{\bR}$ by
\begin{align}
\ell_G(x,Q) \triangleq \langle G'(Q),Q\rangle - G(Q) - G_x'(Q),
\end{align}
where $G'(Q)$ denotes a sub-gradient of $G(Q)$ at $Q$, and $G'_x(Q)$ is the component of $G'(Q)$ corresponding to $Q(x)$ (see, e.g.,  \cite{Gneiting--Raftery2007} for details). The loss function $\ell_G(x,Q)$ also satisfies
\begin{equation}
P \in \inf_{Q \in \Gamma_n} \bE_P[\ell_G(X,Q)]. 
\end{equation}

Substituting loss function $\ell_G(x,Q)$ into \eqref{eqn.overallben} defines a valid $C(\ell, P_{XY})$. The proof is completed via noting that the only non-trivial statistically sufficient transform on a binary alphabet is permutation transform, and the function $G$ is assumed to be invariant to permutations. 

\section{Acknowledgment}

We would like to thank the Associate Editor, two anonymous reviewers, and Yanjun Han, whose comments have significantly improved the presentation of the paper.

\bibliographystyle{IEEEtran}
\bibliography{di}
\begin{IEEEbiographynophoto}{Jiantao Jiao}
(S'13) received his B.Eng. degree with the highest honor in Electronic Engineering from Tsinghua University, Beijing, China in 2012, and a Master's degree in Electrical Engineering from Stanford University in 2014. He is currently working towards the Ph.D. degree in the Department of Electrical Engineering at Stanford University. He is a recipient of the Stanford Graduate Fellowship (SGF). His research interests include information theory and statistical signal processing, with applications in communication, control,
computation, networking, data compression, and learning. 
\end{IEEEbiographynophoto}

\begin{IEEEbiographynophoto}{Thomas A. Courtade}
(S'06-M'13) is an Assistant Professor in the Department of Electrical
Engineering and Computer Sciences at the University of California, Berkeley.
Prior to joining UC Berkeley in 2014, he was a postdoctoral fellow supported
by the NSF Center for Science of Information. He received his Ph.D. and M.S.
degrees from UCLA in 2012 and 2008, respectively, and he graduated summa
cum laude with a B.Sc. in Electrical Engineering from Michigan Technological
University in 2007.

His honors include a Distinguished Ph.D. Dissertation award and an
Excellence in Teaching award from the UCLA Department of Electrical
Engineering, and a Jack Keil Wolf Student Paper Award for the 2012
International Symposium on Information Theory. 
\end{IEEEbiographynophoto}

\begin{IEEEbiographynophoto}{Kartik Venkat}
(S'12) is a Ph.D. candidate in the Department of Electrical
Engineering at Stanford University. His research interests include statistical inference, information theory, machine learning, and their applications in genomics, wireless networks, neuroscience, and quantitative finance. Kartik received a Bachelors degree in Electrical Engineering from the Indian Institute of Technology, Kanpur
in 2010, and a Master's degree in Electrical Engineering from Stanford University in 2012. His honors include a Stanford Graduate Fellowship for Engineering and Sciences, the Numerical Technologies Founders Prize, and a Jack Keil Wolf ISIT Student Paper Award at the 2012 International Symposium on Information Theory.
\end{IEEEbiographynophoto}

\begin{IEEEbiographynophoto}{Tsachy Weissman}
(S'99-M'02-SM'07-F'13) graduated summa cum laude with a B.Sc. in electrical engineering from the Technion in 1997, and earned his Ph.D. at the same place in 2001. He then worked at Hewlett Packard Laboratories with the information theory group until 2003, when he joined Stanford University, where he is currently Professor of Electrical Engineering and incumbent of the STMicroelectronics chair in the School of Engineering. He has spent leaves at the Technion, and at ETH Zurich. 

Tsachy’s research is focused on information theory, compression, communication, statistical signal processing, the interplay between them, and their applications. He is recipient of several best paper awards, and prizes for excellence in research and teaching. He served on the editorial board of the \textsc{IEEE Transactions on Information Theory} from Sept. 2010 to Aug. 2013, and currently serves on the editorial board of Foundations and Trends in Communications and Information Theory. He is Founding Director of the Stanford Compression Forum.  
\end{IEEEbiographynophoto}

\appendix

\section{Proof of Lemmas}

%
%
%
%
%
%

\subsection{Proof of Lemma~\ref{lemma.jiao}}

It follows from (\ref{eqn.vlessthaninfty}) that if we define 
\begin{equation}\label{eqn.defvapp}
V(P) = - \inf_{\hat{x} \in \hat{\mathcal{X}}}\bE_P [\ell(X,\hat{x})],
\end{equation}
then $V(P)$ cannot take values in $\{\infty,-\infty\}$.  

Since $\bE_P [\ell(X,\hat{x})]$ is linear in $P$, and $V(P)$ is the pointwise supremum over a family of linear functions of $P$, we know $V(P)$ is convex and lower semi-continuous on $\Gamma_n$. 

Since $\Gamma_n$ is a compact set, we know that the lower semi-continuous function $V(P)$ attains its minimum on $\Gamma_n$. 

At the same time, since $\Gamma_n$ is a polytope, we know $\forall P = (p_1,p_2,\ldots,p_n) \in \Gamma_n$, we have $P = \sum_{i = 1}^n p_i \delta_i$, where $\delta_i = (0,0,\ldots,1,0,\ldots,0)$ is a distribution that puts mass one at symbol $i$. 

Since $V(P)$ is convex, we have
\begin{equation}
V(P)  = V(\sum_{i = 1}^n p_i \delta_i) \leq \sum_{i = 1}^n p_i V(\delta_i) \leq \max \{V(\delta_i), 1\leq i\leq n\}.
\end{equation}

That is to say, the function $V(P)$ attains its maximum at one of the boundary points $\delta_i$. Thus, we know that $V(P)$ is bounded. 

Now we proceed to show that
\begin{equation}\label{eqn.directdefinv}
\inf_{\hat{X}(Y)} \bE_P[\ell(X,\hat{X}(Y))] = -\sum_{y} P_Y(y) V(P_{X|Y=y}). 
\end{equation}

%

First, for any estimator $\hat{X}(Y)$, by the law of iterated expectation, we have
\begin{align}
\bE_P[\ell(X, \hat{X}(Y))] & = \bE_P[\bE_P[\ell(X,\hat{X}(Y))|Y]]  \\
& \geq \bE_P[-V(P_{X|Y=y})] \\
&= -\sum_{y} P_Y(y) V(P_{X|Y=y}). 
\end{align}

Hence,
\begin{equation}\label{eqn.directdefinine}
 \inf_{\hat{X}(Y)} \bE_P[\ell(X,\hat{X}(Y))] \geq -\sum_{y} P_Y(y) V(P_{X|Y=y}). 
\end{equation}

Second, by the definition of infimum, for any $\epsilon>0$, there exists an estimator $\hat{x}_\epsilon(y) \subset \hat{\mathcal{X}}$ such that
\begin{align}
-V(P_{X|Y=y}) > \sum_{x \in \cX} P_{X|Y=y}(x) \ell(x,\hat{x}_\epsilon(y)) - \epsilon. 
\end{align}

Now define an estimator $\hat{X}(Y) = \hat{x}_\epsilon(Y)$. We have
\begin{align}
\bE_P[\ell(X, \hat{X}(Y))] & = \bE_P[\bE_P[\ell(X,\hat{X}(Y))|Y]] \\
& = \bE_P [\bE_P[\ell(X,\hat{x}_\epsilon(Y))|Y] ] \\
& < \bE_P [-V(P_{X|Y=y})+ \epsilon] \\
& =   -\sum_{y} P_Y(y) V(P_{X|Y=y}) +\epsilon. 
\end{align}

By the arbitrariness of $\epsilon$ we have
\begin{align}
\inf_{\hat{X}(Y)} \bE_P[\ell(X,\hat{X}(Y))] \leq -\sum_{y} P_Y(y) V(P_{X|Y=y}). 
\end{align}

Combining it with (\ref{eqn.directdefinine}), we know that (\ref{eqn.directdefinv}) holds. The claim follows from plugging~(\ref{eqn.defvapp}) and (\ref{eqn.directdefinv}) into the definition of $C(\ell,P_{XY})$. 

\subsection{Proof of Lemma~\ref{lemma.symmetric}}

By Lemma~\ref{lemma.jiao}, we know there exists a convex function $V: \Gamma_n \to \bR$, such that
\begin{align}
C(\ell, P_{XY}) = \left( \sum_{y} P_{Y}(y) V(P_{X|Y=y}) \right) - V(P_{X}).
\end{align}

Let $\delta_i \triangleq (0,0,\ldots,1,\ldots,0)$ be a distribution in $\Gamma_n$ that puts mass one on the $i$-th symbol of $\cX$. Define $a_i \triangleq V(\delta_i)$. We know that $a_i \in \bR, \forall i = 1,2,\ldots,n.$

Define the convex function $G: \Gamma_n \to \bR$ as
\begin{equation}
G(P) = V(P) - \sum_{i = 1}^n a_i p_i.
\end{equation}

Now it is easy to verify that $G(\delta_i) = 0, \forall i = 1,2,\ldots,n$. After some algebra we can show that
\begin{align}
C(\ell, P_{XY})  = \left( \sum_{y} P_{Y}(y) G(P_{X|Y=y}) \right) - G(P_{X}).
\end{align}

Taking $Y \in \cX$, and $P_{Y} = (p_1,p_2,\ldots,p_n)$ to be an arbitrary probability distribution. Setting $P_{X|Y=y} = \delta_{y}$, then we have
\begin{equation}
C(\ell, P_{XY}) =  - G(P_{X}) = -G((p_1,p_2,\ldots,p_n)).
\end{equation}

Define $T = \pi(X)$ to be a permutation of $X$, which is sufficient for $Y$. The Data Processing Axiom implies that
\begin{equation}
C(\ell, P_{XY}) = C(\ell, P_{TY}),
\end{equation}

By construction, we have
\begin{equation}
C(\ell, P_{XY})  = -G((p_1,p_2,\ldots,p_n)),
\end{equation}
\begin{equation}
 C(\ell, P_{TY}) = -G((p_{\pi^{-1}(1)},p_{\pi^{-1}(2)}, \ldots, p_{\pi^{-1}(n)} )),
\end{equation}
which implies that the function $G$ is invariant to permutations. We take $c = -(a_1,a_2,\ldots,a_n)$ to finish the proof.

\end{document}